\documentclass[a4paper,11pt]{article}
\usepackage{pos}

\renewcommand{\vec}{\boldsymbol}    % document specific

\newcommand{\smatrix}{\begin{pmatrix}}
\newcommand{\cmatrix}{\end{pmatrix}}

\usepackage{braket}
\title{Bayesian interpretation of Backus-Gilbert methods}
\author[a]{Luigi Del Debbio}
\author*[b]{Alessandro Lupo}
\author[c]{Marco Panero}
\author[d]{Nazario Tantalo}

\affiliation[a]{Higgs Centre for Theoretical Physics, School of Physics and Astronomy, The University of Edinburgh, Peter Guthrie Tait Road, Edinburgh EH9 3FD, UK}

\affiliation[b]{Aix-Marseille Université, Université de Toulon, CNRS, CPT, Marseille, France}

\affiliation[c]{Department of Physics, University of Turin \& INFN, Turin\\
Via Pietro Giuria 1, I-20125 Turin, Italy}
\affiliation[d]{University and INFN of Roma Tor Vergata\\
	Via della Ricerca Scientifica 1, I-00133, Rome, Italy}

\emailAdd{alessandro.lupo@cpt.univ-mrs.fr}

\abstract{The extraction of spectral densities from Euclidean correlators evaluated on the lattice is an important problem, as these quantities encode physical information on scattering amplitudes, finite-volume spectra, inclusive decay rates, and transport coefficients. In this contribution, we show that the Bayesian approach to this ``inverse'' problem, based on Gaussian processes, can be reformulated in a way that yields a solution equivalent, up to statistical uncertainties, to the one obtained in a Backus-Gilbert approach. After discussing this equivalence, we point out its implications for a reliable determination of spectral densities from lattice simulations.}

\FullConference{The 40th International Symposium on Lattice Field Theory (Lattice 2023)\\
July 31st - August 4th, 2023\\
Fermi National Accelerator Laboratory\\}

\begin{document}
\maketitle

\section{Introduction}
The study of the numerical inversion of the Laplace transform has become popular in the lattice community, due to its importance in determining hadronic observables from simulations of quantum chromodynamics, and gauge theories in general. The problem, however, is a particularly challenging one, and requires a careful treatment to overcome the difficulties related to the inherently limited information that lattice data can provide, due to systematic and statistical uncertainties. Several approaches have been devised~\cite{Hansen:2019idp, Kades:2019wtd, Horak:2021syv, Bergamaschi:2023xzx, Buzzicotti:2023qdv} as a means to provide a stable and reliable solution to the problem, leading to an increase in applications~\cite{Bulava:2019kbi,Gambino:2020crt, Bailas:2020qmv, Bulava:2021fre, Gambino:2022dvu,  ExtendedTwistedMassCollaborationETMC:2022sta, DelDebbio:2022qgu, Pawlowski:2022zhh, Bonanno:2023ljc, Evangelista:2023vtl, Frezzotti:2023nun, Barone:2023tbl}. 

In this work, we focus on two popular methods to tackle this inverse problem: the variation of the Backus-Gilbert (BG) procedure introduced in Ref.~\cite{Hansen:2019idp} and a Bayesian approach based on Gaussian Processes (GP)~\cite{Horak:2021syv, 10.1093/gji/ggz520, DelDebbio:2021whr, Candido:2023nnb}. Even though these two approaches are based on drastically different philosophies we shall prove, by building on the results of Ref.~\cite{10.1093/gji/ggz520} and expanding the discussion present in Ref.~\cite{ExtendedTwistedMassCollaborationETMC:2022sta}, that a suitable choice of the inputs produces a Bayesian solution centred around the modified BG prediction. After a brief introduction in Section~\ref{sec:formulation_of_the_problem} we describe the Bayesian solution to the inversion problem in Section~\ref{sec:Bayesian_Inference_with_Gaussian_Processes} which we then generalise, in Section~\ref{sec:Backus-Gilbert_methods_in_the_Bayesian_framework}, to match the results from Ref.~\cite{Hansen:2019idp}.

\section{Formulation of the problem}
\label{sec:formulation_of_the_problem}

In lattice simulations, Euclidean correlators $C_{LT}$ computed in a finite hypervolume $L^3 \times T$ are related to the spectral density $\rho_{LT}(E)$ via a (generalised) Laplace transform,
\begin{equation}\label{eq:laplace_transform}
    C_{LT}(t) = \int_{0}^\infty dE \; b_T(t,E) \, \rho_{LT}(E) \; ,
\end{equation}
that has to be inverted to extract $\rho_{LT}(E)$. We define $C_L(t)$ as the correlator in the limit $T \rightarrow \infty$, where $b_{T}(t,E) \rightarrow e^{-tE}$. Spectral densities are especially hard to manage in a finite volume, where they are a sum of Dirac $\delta$ distributions across the discrete spectrum of the Hamiltonian. For this reason, numerical methods typically target a smeared version of the spectral density, whereby the finite-volume function $\rho_L(E)$ is convoluted with a Schwartz function $\mathcal{S}_\sigma(\omega,E)$,
\begin{align}\label{eq:smearing}
    & \rho_L(\sigma;\omega) = \int_0^\infty dE \, \mathcal{S}_\sigma(\omega,E) \, \rho_L(E) \, ,\\[8pt]
    & \lim_{\sigma \rightarrow 0} \mathcal{S}_\sigma(\omega,E) = \delta(\omega-E) \; .
\end{align}
Eq.~\eqref{eq:smearing} provides a way to define the infinite-volume limit for the spectral density by the following non-commuting double limit
\begin{equation}
    \rho(\omega) = \lim_{\sigma \rightarrow 0} \lim_{L \rightarrow \infty} \rho_{L}(\sigma;\omega) \; .
\end{equation}
As well documented in the literature, the problem of extracting $\rho_L(\sigma,\omega)$ from $C_L(t)$ is ill-defined. To clarify this point, we begin by noting that if one had access to an infinite set of discrete data that were exact, i.e., unaffected by uncertainties, then the solution could be written as a linear combination of the data
\begin{equation}\label{eq:rho_equals_sum_gt_ct_EXACT}
    \rho^{\rm exact}(\sigma; \omega) = \sum_{\tau=1}^{\infty} g^{\rm exact}_\tau(\sigma;\omega) \, C^{\rm exact}(a \tau) \; .
\end{equation}
It is important to note that the previous infinite sum is an \emph{exact} representation of the continuous smeared spectral density even if the data only constitute a discrete set. A first obstruction arises from the fact that, in reality, the correlator is available only at a finite number of data points $0 < \tau \leq \tau_{\max} < T/a$ (where $\tau = t / a$), which results into a systematic error
\begin{equation}\label{eq:rho_equals_sum_gt_ct_FINITE}
    \rho^{\rm exact}(\sigma ; \omega) = \sum_{\tau =1}^{\tau_{\rm max}} g_\tau(\sigma; \omega) \, C^{\rm exact}(a \tau) + \delta_{\rm sys}(\sigma, \tau_{\rm max}) \; .
\end{equation}
The coefficients $g_\tau(\sigma; \omega)$ from Eq.~\eqref{eq:rho_equals_sum_gt_ct_FINITE} are typically very large and change sign swiftly, a property that is necessary in order to reproduce a smooth function out of the exponentially decaying kernels of the correlators. This leads to a major difficulty in inverting Eq.~\eqref{eq:laplace_transform}. The correlators obtained from lattice simulations are, in fact, unavoidably  affected by statistical and systematic uncertainties, making Eq.~\eqref{eq:rho_equals_sum_gt_ct_FINITE} numerically unstable. A way to tackle this problem consists in “regularising” the sum
\begin{equation}
    \rho(\sigma;\omega) = \sum_{\tau=1}^{\tau_{\rm max}} g_\tau(\sigma;\omega) C(a \tau) 
\end{equation}
by reducing the size of the $g_\tau$ coefficients, so that $\rho(\sigma;\omega)$ does not depend too strongly on the noise of the correlators. At the same time, the “regularised” coefficients should still yield meaningful results with controlled uncertainties. Before describing two of these regularisations in the following section, we remark that any linear combination of correlators necessarily reproduces a smeared spectral density,
\begin{equation}\label{eq:every_reconstruction_is_smeared}
   \sum_{\tau =1}^{\tau_{\rm max}} g_\tau(\sigma;\omega) \, C(a \tau) =   \int dE \left( \sum_{\tau =1}^{\tau_{\rm max}} g_\tau(\sigma;\omega) \,  b_T(a \tau,E)  \right)\, \rho_L(E) \; ,
\end{equation}
the smearing kernel being
\begin{equation}\label{eq:every_reconstruction_is_smeared_PT2}
    \mathcal{S}_\sigma(E,\omega) = \sum_{\tau =1}^{\tau_{\rm max}} g_\tau(\sigma;\omega) \,  b_T(a \tau,E)  \; .
\end{equation}

\section{Bayesian Inference with Gaussian Processes}
\label{sec:Bayesian_Inference_with_Gaussian_Processes}

Let $\mathcal{R}$ be a stochastic field described by the Gaussian Process centred around $\rho^{\rm prior}$ and with covariance $\mathcal{K}^{\rm prior}$
\begin{equation}
    \Pi[\mathcal{R}]  = \frac{1}{\mathcal{N}} \, \exp \biggr( -\frac{1}{2} \left| \mathcal{R} - \rho^{\rm prior} \right|^2_{\mathcal{K}^{\rm prior}} \biggr) \; , 
\end{equation}
where we introduced the norm
\begin{equation}
    \left| \mathcal{R} - \right.
    \left.\rho^{\rm prior} \right|^2_{\mathcal{K}^{\rm prior}} =    \int dE_1 \int dE_2 
    \left[ \mathcal{R}(E_1) - \rho^{\rm prior}(E_1) \right] \mathcal{K}_{\rm prior}^{-1}(E_1,E_2)
     \left[\mathcal{R}(E_2) - \rho^{\rm prior}(E_2) \right] \; ,
\end{equation}
and the normalisation
\begin{equation}
    \mathcal{N} = \int \mathcal{D} \mathcal{R} \; \Pi[\mathcal{R}] \; .
\end{equation}
In the last expressions, $\mathcal{D} R$ is the functional integration measure over the field variable $\mathcal{R}$, which we shall use to describe a continuous function such as the spectral density.

Let $\vec{{\mathcal{C}}} \in \mathbb{R}^{\tau_{\rm max}}$ be a vector of stochastic variables, with components
\begin{equation}\label{eq:laplace_transform_with_noise}
    \mathcal{C}(t) = \int dE \, b_T(t,E) \, \mathcal{R}(E) + \eta(t) \; .
\end{equation}
The vector $\vec{\eta} \in \mathbb{R}^{\tau_{\rm max}}$, which describes the statistical fluctuations of the lattice correlators, is represented as a real-valued stochastic variable, for which we assume a multivariate Gaussian distribution with zero mean and covariance $\text{Cov}_d$,
\begin{equation}
    \mathbb{G}[\vec{\eta}, \text{Cov}_d] = \frac{1}{\sqrt{\text{det} (2\pi \text{Cov}_d)}} \exp \left( -\frac{1}{2}  \vec{\eta}\; \text{Cov}_d^{-1} \;  \vec{\eta}  \right) \, .
\end{equation}
Eq.~\ref{eq:laplace_transform_with_noise} allows the evaluation of the 
covariance:
\begin{equation}\label{eq:CCSigmaHatPlusB}
    %\begin{split}
        \text{Cov}\left[\mathcal{C}(t_1), 
        \mathcal{C}(t_2)\right]  
        %& = \int d\vec{\eta}\;\mathcal{D}\mathcal{R}\; \mathcal{C}(t_1) \, 
        %\mathcal{C}(t_2) \;
        %\mathbb{G}[\vec{\eta}, \text{Cov}_d] \,
        %\Pi[\mathcal{R}] \,  \\[8pt]
        = \Sigma_{t_1t_2} + \left( \text{Cov}_d \right)_{t_1t_2} \; ,
    %\end{split}
\end{equation}
where we defined
\begin{equation}\label{eq:SigmaDef}
    \Sigma_{t_1t_2} = \int dE_1 \int \; dE_2 \; b_T(t_1,E_1) \, \mathcal{K}^{\rm prior}(E_1,E_2) \, b_T(t_2,E_2) \, .
\end{equation}
We are interested in the value of the spectral density at some energy $\omega$, its covariance and its correlation with $\mathcal{C}$; for this purpose we extend the dimensionality of the covariance of Eq.~\eqref{eq:CCSigmaHatPlusB} as follows. Let us introduce the vector $\vec{F} \in \mathbb{R}^{\tau_{\rm max}}$, 
\begin{align}\label{eq:Fvector}
    \text{Cov}\left[\mathcal{C}(t), \mathcal{R}(\omega)\right] 
    %&= 
    %\int d\vec{\eta}\, 
    %\mathcal{D}\mathcal{R} \; \mathcal{C}(t) \, \mathcal{R}(\omega)\,  
    %\mathbb{G}[\vec{\eta}, \text{Cov}_d] \,\Pi[R] \\
    &= F_t(\omega)\;   , 
\end{align}
and the scalar function $F_*$,
\begin{align}
    \label{eq:f_star}
    \text{Cov}\left[\mathcal{R}(\omega), \mathcal{R}(\omega)\right]
    %&= \int \mathcal{D}\mathcal{R} \; \mathcal{R}(\omega)^2
    %\; \Pi[\mathcal{R}]  = \mathcal{K}^{\rm prior}(\omega,\omega) \\
    &= F_*(\omega)\; .
\end{align}
The extended covariance is then 
\begin{equation}\label{eq:total_covariance}
    \Sigma^{\rm tot} = \begin{pmatrix} F_* & \vec{F}^T \\ \vec{F} & \Sigma+\text{Cov}_d
    \end{pmatrix} \; .
\end{equation}
Let $C^{\rm obs}(t)$ be the central value of the correlator measured on the lattice by averaging over a gauge ensemble, and let us denote $\vec{C}^{\, \rm prior}$ the vector of components
%\footnote{Note that Eq.~\eqref{eq:cprior} could be also written in terms of the stochastic variable $\mathcal{C}(t)$, since $\mathcal{C}^{\rm prior}(t) = C^{\rm prior}(t)$.}
\begin{equation}\label{eq:cprior}
    C^{\rm prior}(t) = \int dE \, b_T(t,E) \, \rho^{\rm prior}(E) \; .
\end{equation}
The joint probability density for $\mathcal{R}(\omega)$ and $\mathcal{C}(t)$ is the multidimensional Gaussian
\begin{equation}
    \mathbb{G} \left[\mathcal{R} - \rho^{\rm prior} , \, \vec{\mathcal{C}} - \vec{C}^{\; \rm prior} ; \, \Sigma^{\rm tot} \right] \; ,
\end{equation}
which can be factorised into the product of two Gaussian distributions by performing a block diagonalisation of the total covariance in Eq.~\eqref{eq:total_covariance}. Of the two resulting distributions, one represents the posterior probability density for $\mathcal{R}(\omega)$ given its prior distribution and the set of measurements for the correlator, while the other is the likelihood of the data:
\begin{equation}\label{eq:conditional_probability_factorised}
    \mathbb{G} \left[\mathcal{R} - \rho^{\rm prior} , \, \vec{\mathcal{C}} - \vec{C}^{\; \rm prior} ; \, \Sigma^{\rm tot} \right] = \mathbb{G} \left[ \mathcal{R}- \rho^{\rm post}; \, \mathcal{K}^{\rm post}\right]  \; \mathbb{G} \left[ \vec{\mathcal{C}} - \vec{C}^{\; \rm prior} ; \, \Sigma + \text{Cov}_d \right] \; .
\end{equation}
The posterior Gaussian distribution for the spectral density is centred around:
\begin{equation}\label{eq:naive_gp_mean}
    \left. \rho^{\rm post}(\omega) \right|_{\mathcal{C}=C^{\rm obs}} = \rho^{\rm prior}(\omega) + \vec{F}^T  \frac{1}{\Sigma + \text{Cov}_d}   \left( \vec{C}^{\;\rm obs} - \vec{C}^{\; \rm prior} \right) \; ,
\end{equation}
and has variance
\begin{equation}\label{eq:naive_gp_variance}
    \left. \mathcal{K}^{\rm post}(\omega,\omega)\right|_{\mathcal{C}=C^{\rm obs}} =
     \mathcal{K}^{\rm prior}(\omega,\omega) - \vec{F}^T  \frac{1}{\Sigma + \text{Cov}_d}  \vec{F} \; .
\end{equation}
In order to make contact with Eq.~\eqref{eq:rho_equals_sum_gt_ct_FINITE}, we introduce the coefficients
\begin{equation}\label{eq:gt_naive_GP}
    \vec{g}^{\rm \, GP} (\omega) =  \vec{F}^T  \frac{1}{\Sigma + \text{Cov}_d} \; .
\end{equation}
Let us discuss the previous equations. Eq.~\eqref{eq:total_covariance} shows how the inverse problem is regularised once expressed in terms of probability distributions: the numerical instability, which leads to the very large coefficients of Eq.~\eqref{eq:rho_equals_sum_gt_ct_EXACT}, is due to the ill-conditioning of the model covariance $\Sigma$ in Eq.~\eqref{eq:SigmaDef}. The covariance $\text{Cov}_d$, added to the matrix $\Sigma$, cuts off its near-zero eigenvalues. The resulting solution is then stable within statistical uncertainties. The very same regularisation is used in Backus-Gilbert methods~\cite{10.1111/j.1365-246X.1968.tb00216.x, Hansen:2019idp, ExtendedTwistedMassCollaborationETMC:2022sta}, despite the fact that they contain no formulation in terms of stochastic variables.

As pointed out around Eqs.~\eqref{eq:every_reconstruction_is_smeared} and~\eqref{eq:every_reconstruction_is_smeared_PT2}, the centre of the posterior describes a smeared spectral density, even if the smearing was not one of the initial assumptions. The smearing function, which can be inferred from Eq.~\eqref{eq:every_reconstruction_is_smeared_PT2}, is determined by the choice of the priors and the noise on the data, as seen in Eq.~\eqref{eq:gt_naive_GP}. Consequently, the smearing kernel obtained in this setup is not known a priori. In this regard, the solution given in this section is similar to the original Backus-Gilbert proposal~\cite{10.1111/j.1365-246X.1968.tb00216.x} rather than the method of Ref.~\cite{Hansen:2019idp}, where the smearing kernel is chosen. The direct connection with Ref.~\cite{Hansen:2019idp} is given in the next section.

\section{Backus-Gilbert methods in the Bayesian framework}
\label{sec:Backus-Gilbert_methods_in_the_Bayesian_framework}
The first step is to target the probability density for a spectral density smeared with an input kernel $\mathcal{S}_\sigma(\omega,E)$. To generalise the results of the previous section, we introduce the stochastic variable 
\begin{equation}
    \mathcal{R}_\sigma(\omega) = \int dE \; \mathcal{S}_\sigma(\omega,E) \, \mathcal{R}(E) \; .
\end{equation}
The same steps described in Section~\ref{sec:Bayesian_Inference_with_Gaussian_Processes} lead to the following extended covariance,
\begin{equation}
\Sigma^{\rm tot} = \begin{pmatrix}
F^{\sigma *} & \vec{F}^{\sigma} \\ \vec{F}^{\sigma} & \Sigma + \text{Cov}_d
\end{pmatrix} \; ,
\end{equation}
where $\Sigma$ is as in Eq.~\eqref{eq:SigmaDef}, while the other functions now include reference to the smearing kernel:
\begin{align}
F^{\sigma}_*(\omega) &= \int dE_1 \int dE_2 \; \mathcal{S}_\sigma(\omega,E_1) \, \mathcal{K}^{\rm prior}(E_1, E_2) \, \mathcal{S}_\sigma(E_2, \omega) \; , \\
F^{\sigma}_{t}(\omega) &= \int dE_1 \int dE_2 \;  \, b_T(a\tau, E_1) \, \mathcal{K}^{\rm prior}(E_1, E_2) \, \mathcal{S}_\sigma(E_2,\omega) \, . 
\end{align}
To match Ref.~\cite{Hansen:2019idp} one can select~\cite{ExtendedTwistedMassCollaborationETMC:2022sta} a diagonal model covariance,
\begin{equation}\label{eq:hltgp_prior}
\mathcal{K}^{\rm prior}(E_1, E_2) = \frac{e^{\alpha E}}{\lambda} \delta(E_1-E_2) \; ,
\end{equation}
where $\alpha < 2$ and $\lambda \in (0,\infty)$ are, in this context, hyperparameters, which can be chosen by maximising the data likelihood~\cite{Horak:2021syv}. They are introduced to match the parameters appearing in Eq.~(A3) in Ref.~\cite{Hansen:2019idp}. 

Let the smearing kernel be, for instance, a Gaussian $G_\sigma(\omega,E) = \exp[-(\omega-E)^2/2\sigma^2] / \sqrt{2\pi}\sigma$. The posterior Gaussian distribution for a smeared spectral density, given the prior distribution and the observed data, is centred around
\begin{equation}\label{eq:hltgp_central}
\rho^{\rm post}_\sigma(\omega) =  \rho^{\rm prior}_\sigma(\omega) + \sum_{\tau=1}^{\tau_{\rm max}}  g^{\rm BG}_\tau(\sigma, \omega) \, C(a \tau) \; ,
\end{equation}
and has variance
\begin{equation}\label{eq:hltgp_variance}
\mathcal{K}^{\rm post}(\omega,\omega)  =  \left(\int dE \, G^2_\sigma(\omega,E) \, \frac{e^{\alpha \omega}}{\lambda} \right) - \sum_{\tau=1}^{\tau_{\rm max}}  g^{\rm BG}_\tau(\sigma, \omega) F^\sigma_\tau(\omega) \; .
\end{equation}
Due to the careful choice of model covariance in Eq.~\eqref{eq:hltgp_prior}, the coefficients $g^{\rm BG}_\tau(\sigma, \omega)$ are the same that were derived in Ref.~\cite{Hansen:2019idp}. We recall that in the latter, the coefficients $g^{\rm BG}_\tau$ are determined in a drastically different way, i.e. by minimising the following functional:
\begin{multline}\label{eq:A_funcitonal}
    (1 - \lambda') \int_{0}^\infty dE \, e^{\alpha E} \left| \sum_{\tau=1}^{\tau_{\rm max}} \, g_\tau(\sigma;\omega) \, b_T(a\tau,E) - G_\sigma(\omega,E) \right|^2 \\ + \lambda' \sum_{\tau, \tau'=1}^{\tau_{\rm max}} g_\tau(\sigma;\omega) \; \text{Cov}_{\tau \tau'} \; g_{\tau'}(\sigma;\omega) \; . \hspace{1cm} 
\end{multline}
The Backus-Gilbert parameter $\lambda' \in (0,1)$ is related to the $\lambda$ in Eq.~\eqref{eq:hltgp_prior} by $\lambda = \lambda'/(1-\lambda')$.

We identified a setup in which Bayesian and Ref.~\cite{Hansen:2019idp}, two frameworks with utterly different philosophies, provide the same central value for the smeared spectral densities. There are, however, important differences between the two methods, that we shall now discuss. A first aspect concerns the error on the smeared spectral density. In non-Bayesian methods, including the one introduced in Ref.~\cite{Hansen:2019idp}, statistical uncertainties are often propagated from the data by bootstrap. In the context of GPs, on the other hand, by working with Gaussian distributions we are able to provide an analytic expression for the error, which is inherited by Eq.~\eqref{eq:hltgp_variance}, the variance of the probability density that describes the smeared spectral densities. Another important difference is the way algorithmic parameters are determined. For the Backus-Gilbert method, a procedure that has been proven effective~\cite{ExtendedTwistedMassCollaborationETMC:2022sta, Evangelista:2023fmt, DelDebbio:2022qgu} was introduced in Ref.~\cite{Bulava:2021fre}: it consists in finding a range of parameters, $\lambda$ and $\alpha$, such that any shift in the smeared spectral density is smaller than statistical fluctuations. In this way, one ensures that the result does not depend on unphysical parameters of the algorithm. In Bayesian inference, $\lambda$ and $\alpha$ have a different interpretation as they are hyperparameters that determine the prior distribution. The latter, however, are again chosen \emph{ad hoc} as inputs of the procedure, hence they should not affect the final result (up to statistical fluctuations). In Bayesian inference, it is common to determine the hyperparameters by minimising the negative logarithmic likelihood (NLL),
\begin{multline}\label{eq:NLL}
\frac{\tau_{\rm max}}{2}\, \text{Log} (2\pi) +\frac{1}{2} \, \text{Log}\, \text{det} \left( \Sigma + \text{Cov}_d \right) +\frac{1}{2} \, (\vec{C}^{\, \rm obs}- \vec{C}^{\; \rm prior}) \frac{1}{\Sigma+\text{Cov}_d}  (\vec{C}^{\, \rm obs}- \vec{C}^{\; \rm prior}) \; .
\end{multline}

\begin{figure}[tb!]
    \centering
    \includegraphics[width=0.6\columnwidth]{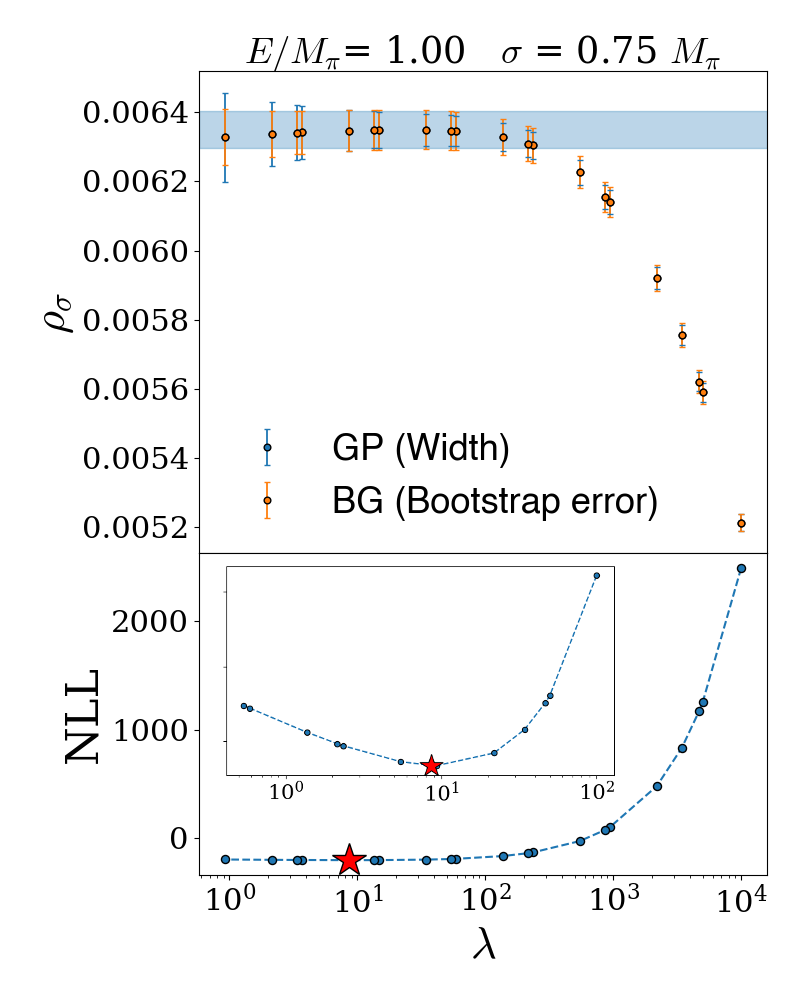}
    \caption{Top panel: spectral density smeared with a Gaussian, for different values of the $\lambda$-parameter. The central value is obtained according to Ref.~\cite{Hansen:2019idp}, or equivalently from Eq.~\eqref{eq:hltgp_central}. The error is computed with a bootstrap in the former case (BG in the legend), and is given by Eq.~\eqref{eq:hltgp_variance} in the latter (GP in the legend). The figure also shows the stability region described in Ref.~\cite{Bulava:2021fre}, which predicts a value identified by the horizontal band. This is consistent with the value obtained by choosing the $\lambda$ that minimises the NLL (red star). The results are obtained using lattice data from Ref.~\cite{DelDebbio:2022qgu}.}
    \label{fig:gphlt-scan}
\end{figure}

In light of the analogy described in this work, it is natural to ask how values of $\{\lambda, \alpha\}$ specified by the minimum of the NLL relate to values determined according to the non-Bayesian procedure of Refs.~\cite{Hansen:2019idp, Bulava:2021fre}. To this end we show, in the top panel of Fig.~\ref{fig:gphlt-scan}, a comparison between a smeared spectral density obtained in the two approaches, as a function of $\lambda$. The central values are the same, as inferred from Eq.~\eqref{eq:hltgp_central}. The statistical errors are determined with a bootstrap procedure (BG in the legend) or from the square root of half Eq.~\eqref{eq:hltgp_variance} (GP in the legend). In this example, which uses Monte Carlo lattice data\footnote{The correlator used is a two-point correlation function of pseudoscalar mesons, in the two-index antisymmetric representation of $SU(4)$ gauge theory, corresponding to the ensemble $B3$ of Ref.~\cite{DelDebbio:2022qgu}.} from Ref.~\cite{DelDebbio:2022qgu}, the uncertainties are found to be of the same order of magnitude, which is a non-trivial result, given the profound difference in the way they are obtained. Another striking observation is that the value of $\lambda$ determined from the minimum of the NLL (red star in the bottom panel), lies in a region in which the smeared spectral density does not change, within statistical noise, by changing the value of $\lambda$. The value for the smeared spectral density that one would obtain following Ref.~\cite{Bulava:2021fre}, shown as a horizontal band in the top panel of Fig.~\ref{fig:gphlt-scan}, is in fact consistent with the value obtained minimising the NLL. This provides a non-trivial validation for both methods and, eventually, for the predictions they yield.

\section{Conclusions}
\label{sec:Conclusions}

The method described in Ref.~\cite{Hansen:2019idp} and Gaussian processes are popular choices to compute smeared spectral densities from lattice correlators. We have shown that the approach proposed in Ref.~\cite{Hansen:2019idp} can be reformulated in a Bayesian framework. This leads to a drastically different interpretation of the variables at hand, which become stochastic variables characterised by their probability distributions. The solution, and its error, are understood as the central value and the width of probability distributions. Nonetheless, for a specific choice of priors, the final prediction is identical, within statistical uncertainties, to the one from Ref.~\cite{Hansen:2019idp}. In our numerical tests, based on lattice correlation functions of meson-like states, we have found that the Bayesian error on the spectral reconstruction, coming from the width of its probability density, is compatible with the statistical error that one obtains by bootstrapping in a frequentist fashion. The analogy extends to the determination of the input parameters, where the prescription of Ref.~\cite{Bulava:2021fre} for the $\lambda$ parameter is found to be compatible with the minimisation of the NLL. Further details on this comparison will be presented in a forthcoming publication.

\section*{Acknowledgments}
AL is funded in part by l’Agence Nationale de la Recherche (ANR), under grant ANR-22-CE31-0011. AL and LDD received funding from the European Research Council (ERC) under the European Union’s Horizon 2020 research and innovation program under Grant Agreement No.~813942. LDD is also supported by the UK Science and Technology Facility Council (STFC) grant ST/P000630/1. MP was partially supported by the Spoke 1 ``FutureHPC \& BigData'' of the Italian Research Center on High-Performance Computing, Big Data and Quantum Computing (ICSC) funded by MUR (M4C2-19) -- Next Generation EU (NGEU), by the Italian PRIN ``Progetti di Ricerca di Rilevante Interesse Nazionale -- Bando 2022'', prot. 2022TJFCYB, and by the ``Simons Collaboration on Confinement and QCD Strings'' funded by the Simons Foundation.

\newpage
\bibliographystyle{unsrt}         
\bibliography{main}
\end{document}